\documentclass[letterpaper]{article} % DO NOT CHANGE THIS
\usepackage{aaai23}  % DO NOT CHANGE THIS
\usepackage{times}  % DO NOT CHANGE THIS
\usepackage{helvet}  % DO NOT CHANGE THIS
\usepackage{courier}  % DO NOT CHANGE THIS
\usepackage[hyphens]{url}  % DO NOT CHANGE THIS
\usepackage{graphicx} % DO NOT CHANGE THIS
\usepackage{natbib}  % DO NOT CHANGE THIS AND DO NOT ADD ANY OPTIONS TO IT
\usepackage{caption} % DO NOT CHANGE THIS AND DO NOT ADD ANY OPTIONS TO IT
\usepackage{algorithm}
\usepackage{algorithmic}

\usepackage{newfloat}
\usepackage{listings}
\DeclareCaptionStyle{ruled}{labelfont=normalfont,labelsep=colon,strut=off} % DO NOT CHANGE THIS
\floatstyle{ruled}
\newfloat{listing}{tb}{lst}{}
\floatname{listing}{Listing}
\usepackage{amsmath}
\usepackage{amssymb}
\title{Self-Supervised Learning for Modeling Gamma-ray Variability in Blazars}
\author {
    Aryeh Brill
}
\affiliations{
    NASA Postdoctoral Program Fellow, NASA Goddard Space Flight Center\\
    aryeh.brill@nasa.gov
}

\begin{document}
\nocopyright

\maketitle

\begin{abstract}
Blazars are active galactic nuclei with relativistic jets pointed almost directly at Earth. Blazars are characterized by strong, apparently stochastic flux variability at virtually all observed wavelengths and timescales, from minutes to years, the physical origin of which is still poorly understood. In the high-energy gamma-ray band, the Large Area Telescope aboard the \textit{Fermi} space telescope (\textit{Fermi}-LAT) has conducted regular monitoring of thousands of blazars since 2008. Deep learning can help uncover structure in gamma-ray blazars' complex variability patterns that traditional methods based on parametric statistical modeling or manual feature engineering may miss. In this work, we propose using a self-supervised Transformer encoder architecture to construct an effective representation of blazar gamma-ray variability. Measurement errors, upper limits, and missing data are accommodated using learned encodings. The model predicts a set of quantiles for the flux probability distribution at each time step, an architecture naturally suited for describing data generated by a stochastic process. As a proof of concept for how the model output can be analyzed to extract scientifically relevant information, a preliminary search for weekly-timescale time-reversal asymmetry in gamma-ray blazar light curves was conducted, finding no significant evidence for asymmetry. 

\end{abstract}

\section{Introduction}

Blazars are the most extreme class of active galactic nuclei (AGN), energetic phenomena powered by accretion onto supermassive black holes in the centers of a few percent of galaxies. A fraction of AGN host relativistic jets that accelerate particles close to the speed of light. AGN with jets oriented close to the line of sight towards Earth are called blazars \citep{Urry1995}. Blazars are the most luminous long-lasting sources of electromagnetic radiation in the gamma-ray sky. Blazar emission extends across the electromagnetic spectrum, including gamma rays in the high-energy (HE; $\sim$0.1-100 GeV) band and even higher energies.

HE gamma-ray observations of blazars have been conducted since 2008 by the Large Area Telescope on board the \textit{Fermi} Gamma-Ray Space Telescope (\textit{Fermi}-LAT). \textit{Fermi}-LAT detects gamma rays from 20 MeV to above 500 GeV using a pair-conversion technique \citep{Atwood2009}. Located in low Earth orbit, \textit{Fermi}-LAT primarily operates in survey mode, during which it scans the entire sky every 3 hr. Among other data products, \textit{Fermi}-LAT generates binned gamma-ray flux light curves (time series) for each detected source. The time bin duration (typically chosen between 1 day and several months) is a trade-off, with a shorter duration improving time resolution but worsening sensitivity.

Blazars can be divided into two main classes based on their optical spectra, flat spectrum radio quasars (FSRQs) and BL Lacertae-type objects (BL Lacs). The brightest and most variable objects in HE gamma rays are typically FSRQs. Over 3000 blazars have been detected by \textit{Fermi}-LAT \citep{Ajello2022}. Of the \textit{Fermi}-LAT blazars, 755 are FSRQs, 1379 are BL Lacs, and 1208 are of an unknown type.

The physical processes that cause gamma-ray blazar variability are poorly understood. Long-timescale ($\gtrsim 1$~year) gamma-ray variability may be connected to processes in the accretion disk, while short-timescale ($\lesssim 1$~day) variability may result from emission occurring in compact structures in the jet \citep{Rieger2019}. Some gamma-ray blazars exhibit apparent trends \citep[e.g.][]{Valverde2020} or (quasi-)periodic oscillations \citep[e.g.][]{Penil2020, Rueda2022} on timescales of years. 

Of particular physical and observational interest is intermediate-timescale variability, ranging from days to months. Most strikingly, on these timescales blazars can undergo flares, or short-lived flux increases of as much as two orders of magnitude \citep[e.g.][]{Adams2022}. The physical origin of gamma-ray flares in blazars, or even whether flares should be understood as a physical process distinct from ordinary gamma-ray emission, is not known. For example, a model of intermediate-timescale variability as the interaction between an unresolved short-timescale burst process and long-timescale stochastic variations can yield flares as an emergent property \citep{Brill2022}. An understanding of to what extent flares are asymmetric in time is also valuable, as their rise and decay timescales may be connected to the timescales of the particle acceleration and cooling processes in the emission region \citep[e.g.][]{Abdo2010}. For these reasons, a better understanding of blazar variability at intermediate timescales, including the shape of the flux probability distribution and whether it changes over time, can give crucial insight into physical processes.

Blazar emission typically has a power spectral density (PSD) with a power-law shape \citep[e.g.][]{Abdo2010}, possibly with spectral breaks, indicating that it is wholly or partly stochastic. Characteristic gamma-ray variability timescales can be discovered through time-series modeling \citep[e.g.][]{Kelly2009, Ryan2019, Brill2022}. However, variability analysis based on PSD fitting or autoregressive modeling has several important limitations. Model parameters, such as the shape of the flux distribution, must be imposed. Depending on the method used, it can be difficult to deal with missing time bins or upper limits. Methods based on second-order statistics are restricted to modeling time-symmetric autocorrelation. A key assumption underlying many time-series models is stationarity, or time invariance of statistical properties (possibly after filtering out trends or periodicities). The extent to which blazar light curves can be properly described as stationary is still under investigation. For example, \citet{Duda2021} reported transient non-stationarity features in gamma-ray blazar light curves. One potential form of apparent nonstationarity could be a tendency for flaring activity to increase (or suppress) further flux variability, manifesting in time-asymmetric conditional heteroskedasticity. In this work, we explore how the expressive power of deep neural networks can complement traditional analysis tools and enable novel studies by surfacing variability patterns in a model-independent way. 

Several different approaches have been used to analyze gamma-ray blazar variability with deep neural networks. Some works have focused on the supervised task of classifying sources of unknown spectral type as FSRQs or BL Lacs using variability features manually extracted from light curves \citep[e.g.][]{Doert2014, Kaur2019}. In most contexts, however, light curves are unlabeled, making it natural to adopt an unsupervised approach capable of automatically extracting information of scientific interest for further analysis, classification, and interpretation. Furthermore, light curves may contain rich variability patterns that manually determined features do not adequately describe.

One possible approach, increasingly applied in optical astronomy \citep{Huertas-Company2022}, has used recurrent autoencoders in a semi-supervised or anomaly detection framework to search for transient events such as supernovae \citep[e.g.][]{Villar2020, Villar2021} or anomalous variability behavior in AGN \citep{Sanchez-Saez2021}. Using a recurrent autoencoder, \citet{Tachibana2020} found evidence of temporal asymmetry in the optical variability of quasars, though such an asymmetry could have been produced by a selection bias induced at the time of sample selection \cite{Shen2021}. Autoencoders can efficiently represent a light curve's overall long-term variability structure. However, because their compression objective leads them to discard short-term fluctuations, they are less well suited for modeling variability on shorter time scales.

Instead, we propose an approach to uncovering variability patterns in gamma-ray light curves based on self-supervised learning (SSL), in which some of the data points in the input light curves are randomly masked, and a deep neural network is trained to predict the missing data based on its context. We employ a model closely based on Bidirectional Encoder Representations from Transformers \citep[BERT;][]{Devlin2018}, widely used in natural language processing. BERT is built around the Transformer architecture \citep{Vaswani2017}, which employs a self-attention mechanism to efficiently model complex long-term dependencies in data. In astronomy, SSL using Transformers has been used to remove noise and outliers from optical light curves \citep{Morvan2022} and to perform pre-training for classifying the light curves of variable stars \citep{Donoso-Oliva2022}.

In this work, we have applied SSL to extract variability information from gamma-ray light curves of blazars. In order to naturally represent the variability structure of stochastic light curves, we introduce a training objective in which the network predicts a probability distribution, rather than a scalar value, for each missing data point, represented non-parametrically as a set of quantiles. After training, the resulting probability distributions can be analyzed to obtain quantities of scientific interest. As a proof of concept, we applied SSL to search for weekly-timescale time-reversal asymmetry in gamma-ray blazar light curves.

\section{Data}

For our gamma-ray dataset, we used the publicly available light curves from the \textit{Fermi}-LAT Light Curve Repository\footnote{\url{https://fermi.gsfc.nasa.gov/ssc/data/access/lat/LightCurveRepository/}} \citep[LCR;][]{LAT2021}. The LCR consists of light curves for 1525 sources estimated to have a less than 1\% chance of being a steady source over 10 time intervals in the source catalog of the first 10 years of \textit{Fermi}-LAT observations \citep{Ballet2020}. Of these, about 93\% are blazars.

The LCR provides light curves with time bin sizes of 3 days, 7 days, and 30 days. The light curves are generated by a maximum likelihood analysis. For each time bin, the significance is estimated by a likelihood test statistic \citep[$TS$;][]{Mattox1996}. For time bins with a significant detection, the flux $F$ and $1\sigma$ errors $\sigma_F$ are provided; otherwise a 95\% photon flux upper limit $UL$ is reported. Time bins may be missing entirely if insufficient data were collected during a time bin or if the analysis otherwise failed to converge. We downloaded the 7 day photon flux light curves corresponding to a fixed spectral index for all sources for the time interval from August 5, 2008 through March 14, 2022, for a maximum of 710 time bins per source.

Guided by the LCR usage notes\footnote{\url{https://fermi.gsfc.nasa.gov/ssc/data/access/lat/LightCurveRepository/about.html}}, we applied a number of quality cuts to the data. Potentially non-convergent bins were removed by excluding bins with a nonzero return code; $TS \leq 0$; $UL < 0$; $(F/\sigma_F)^2/TS \geq 1$; or $(F/\sigma_F)^2/TS \leq 0.1$. Upper limits were selected for all time bins with $TS < 4$, and low-quality data points with $F \leq \sigma_F$ were excluded.

On March 16, 2018, the \textit{Fermi} spacecraft experienced an anomaly in its drive system, after which it adopted a modified orbital profile\footnote{\url{https://fermi.gsfc.nasa.gov/ssc/observations/types/post_anomaly/}}. This resulted in occasional periods of low exposure for some sources, leading to erroneous photon flux estimates. To remove any data points potentially affected by this anomaly, for each source $i$, data points with $\sigma_F/\mu_i < 1.5$ and upper limits with $UL/2\mu_i < 1.5$ were excluded, where $\mu_i$ is that source's mean photon flux.

\begin{figure}[ht]
    \centering
    \includegraphics[width=0.9\columnwidth]{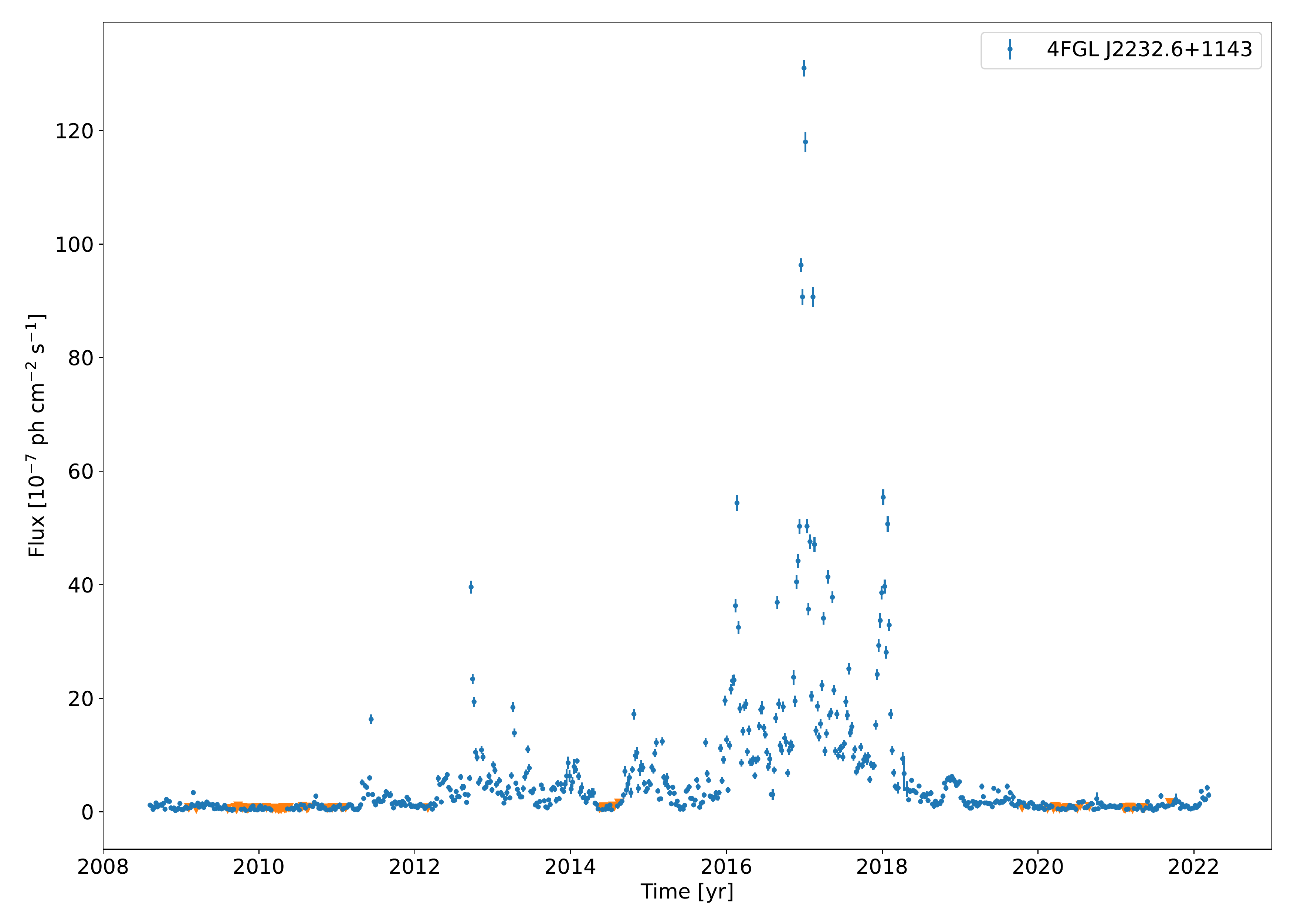}
    \includegraphics[width=0.9\columnwidth]{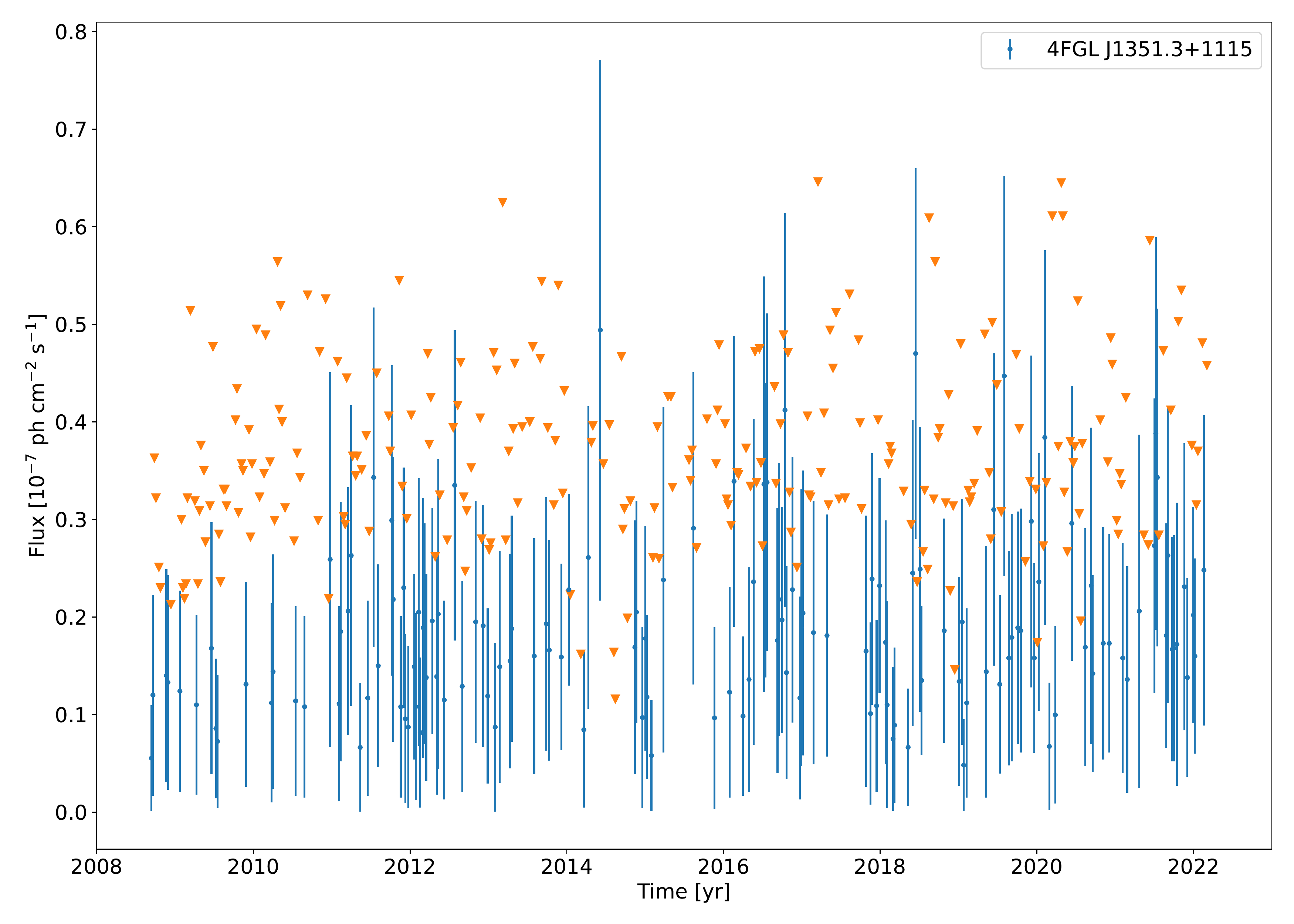}
    \caption{Light curves of the most (top) and least (bottom) variable sources in the LCR dataset after all cuts. Upper limits are indicated by orange triangles.}
    \label{fig:light_curves}
\end{figure}

Finally, any source with fewer than 100 flux data points; fewer than 300 combined data points and upper limits; or not classified as an FSRQ, BL Lac, or blazar candidate of unknown type was excluded, leaving 689 sources. Fig.~\ref{fig:light_curves} shows the light curves after all cuts of the most and least variable sources in the final dataset, as determined by the 4FGL-DR2 variability index \citep{Ballet2020}. Each time bin has either an associated flux $F$ and error $\sigma_F$, or an upper limit $UL$, all in units of photons cm$^{-2}$ s$^{-1}$, as well as an associated time index $0 \leq t < 710$.

\begin{figure*}[t]
    \centering
    \includegraphics[width=0.9\textwidth]{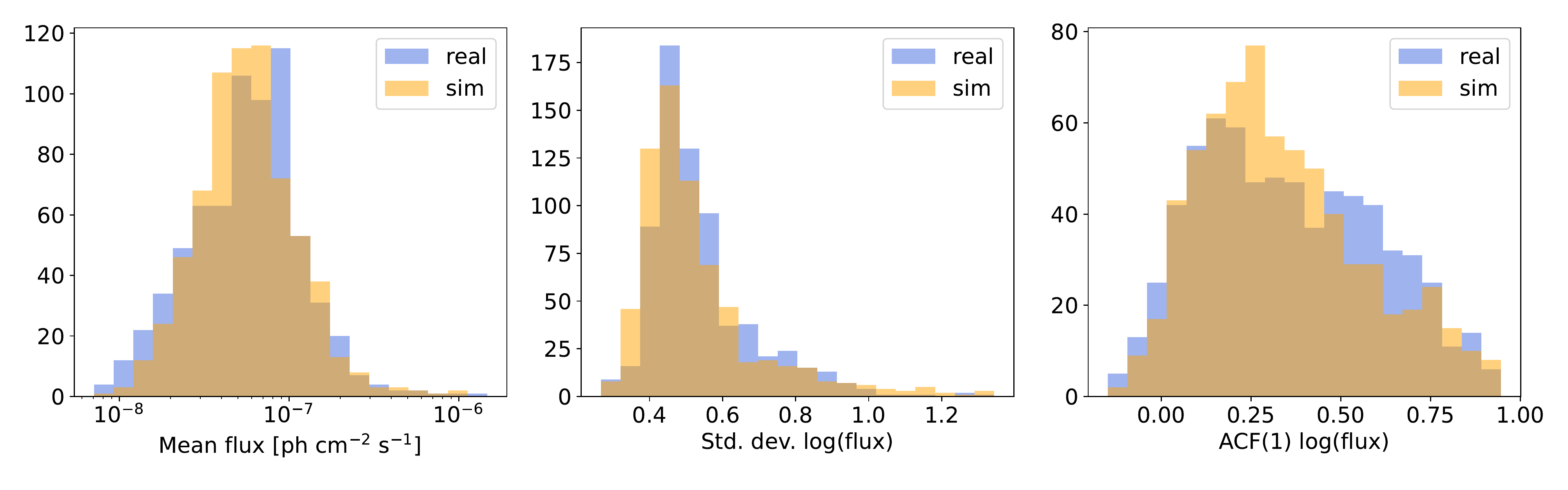}
    \caption{A comparison of statistical properties of the real and simulated datasets, showing, from left to right, the mean photon flux, the standard deviation of the natural logarithm of the flux, and the lag-1 autocorrelation function of the natural logarithm of the flux. Upper limits were excluded from all calculations. No adjustments were made to account for missing time bins.}
    \label{fig:dataset_distributions}
\end{figure*}

In order to develop and validate the network, we also created a simulated dataset intended to roughly emulate the general statistical features of the LCR dataset. 1378 light curves were simulated, of which 689 were used for training and 689 were reserved for validation. Each light curve consisted of 710 time steps and was generated following a first-order autoregressive (AR(1)) process,

\begin{equation}
    y_{t + 1} = \phi y_t + \epsilon_t,
\end{equation}

\noindent where $\epsilon_t \sim \mathcal{N}(0, \sigma)$. The light curves were exponentiated to obtain a lognormal flux distribution and scaled by a random normalization parameter $\mu_N$. The AR(1) parameters of each light curve were randomly selected following the distribution parameters in Table~\ref{tab:lc_parameters}. Simulated measurement errors were applied to the light curves and time bins were randomly removed using a flux/error correspondence and missing-bin probability empirically estimated from the LCR data. Time bins with $F < 2\sigma_F$ were replaced with upper limits such that $UL = \mathrm{max}(2\sigma_F + F, 2\sigma_F)$. A comparison of some statistical properties of the real and simulated datasets is shown in Fig.~\ref{fig:dataset_distributions}.

\begin{table}[ht]
    \centering
    \begin{tabular}{c|c|r|r|r|r}
        Param. & Distribution & $\mu_\mathrm{dist}$ & $\sigma_\mathrm{dist}$ & Min & Max \\
        \hline
        $\phi$ & Trunc. Norm. & 0.70 & 0.30 & 0 & 0.975\\
        $\sigma$ & Trunc. Norm. & 0.45 & 0.05 & 0 & -\\
        $\mu_N$ & Lognormal & -17.4 & 0.8 & - & -\\
    \end{tabular}
    \caption{Distributions of parameters used to generate  the simulated light curves.}
    \label{tab:lc_parameters}
\end{table}

\section{Methodology}

The core of our model consisted of a BERT-style Transformer encoder network with $L = 4$, $H = 32$, and $A = 4$, with $L$ the number of layers, $H$ the hidden size, and $A$ the number of self-attention heads. The feedforward dimension is set to $4H$ and dropout was used with probability 0.1. Adam was used to train the model with a learning rate of $1\times10^{-3}$, $\beta_1 = 0.9$, and $\beta_2 = 0.999$ \citep{Kingma2014}.

We applied several preprocessing steps to transform the data into an input format suitable for the network. First, for each light curve, a segment of 101 time bins was extracted, with the start point chosen randomly. The time indices were shifted to make the index of the central data point 0. Time steps were randomly selected for masking with 20\% probability. Of those selected, a mask was actually applied with 90\% probability, otherwise, the data point was unaltered. The natural logarithm was applied to the flux points, $1\sigma$ flux lower and upper bounds, and flux upper limits. The mean of the unmasked flux points in the segment was subtracted. To create the input vector for the network, the time indices and flux values were converted to vectors of length 32 using fixed trigonometric encodings \citep{Vaswani2017}. Since the flux values and time index both used fixed encodings, the vectors were concatenated rather than summed, yielding input vectors of length 128. For upper limits, the flux upper and lower limits were replaced by a single learned upper limit encoding, while for masked time steps all three flux values were replaced by a single learned mask encoding (superseding any upper limit encoding). This process was repeated in batches, with a batch size of 64.

\begin{figure}[ht]
    \centering
    \includegraphics[width=0.9\columnwidth]{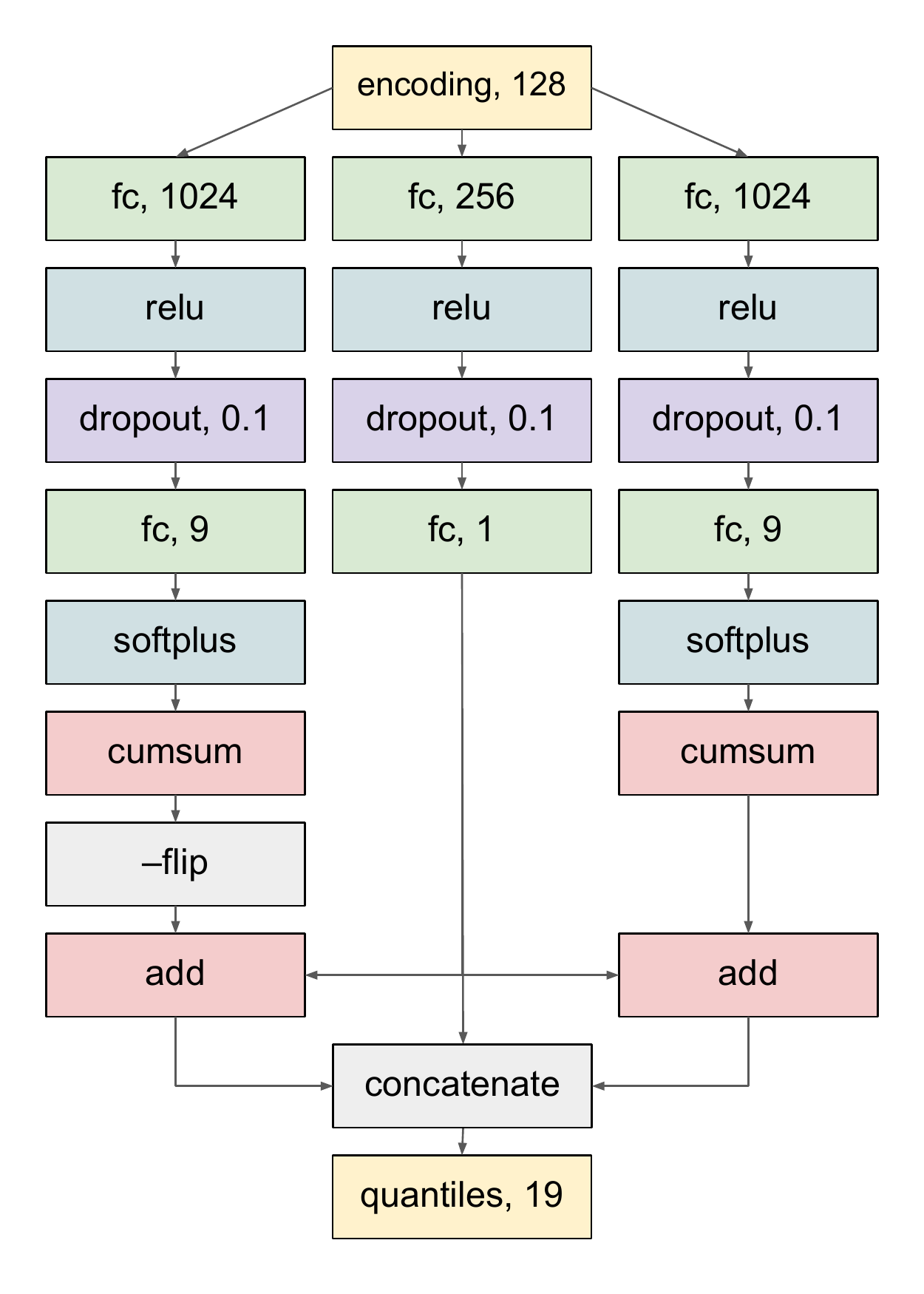}
    \caption{Diagram of the Quantile Head applied to the output embedding at each time step.}
    \label{fig:quantile_head}
\end{figure}

To generate a non-parametric probability distribution for the flux at each time step, the output encoding was fed through a Quantile Head, illustrated in Fig.~\ref{fig:quantile_head}. Three independent stacks of fully-connected networks computed the median; quantiles less than 0.5; and quantiles greater than 0.5, respectively. The Quantile Head efficiently prevents quantile crossing, i.e., non-monotonicity of the estimated quantiles, by applying a softplus activation followed by a cumulative sum to the estimated quantile vectors. The vector for quantiles less than 0.5 was flipped and negated, the median added to the other quantiles, and all of the quantiles finally concatenated. We used the set of 19 quantiles for probabilities 0.05 to 0.95, in increments of 0.05.

The SSL objective was the quantile score or pinball loss function,

\begin{equation}
    \mathcal{L} = 2\sum_p \max(-p(q_p - y), (1 - p)(q_p - y)),
\end{equation}

where $q_p$ is the quantile corresponding to probability $p \in (0, 1)$ and $y$ is the measured value of $\log(F)$. For $p = 0.5$, the quantile score reduces to the absolute loss. Only masked flux points were included in the loss; any masked upper limits were ignored.

\section{Results}

The network was first trained on the simulated dataset for 1000 epochs, monitoring the validation loss throughout training to confirm that it remained consistent with the training loss. The network was then retrained from scratch on the real LCR dataset using the same procedure. Each model was subsequently used to generate predictions on its respective corresponding dataset only.

\begin{figure}[ht]
    \centering
    \includegraphics[width=0.9\columnwidth]{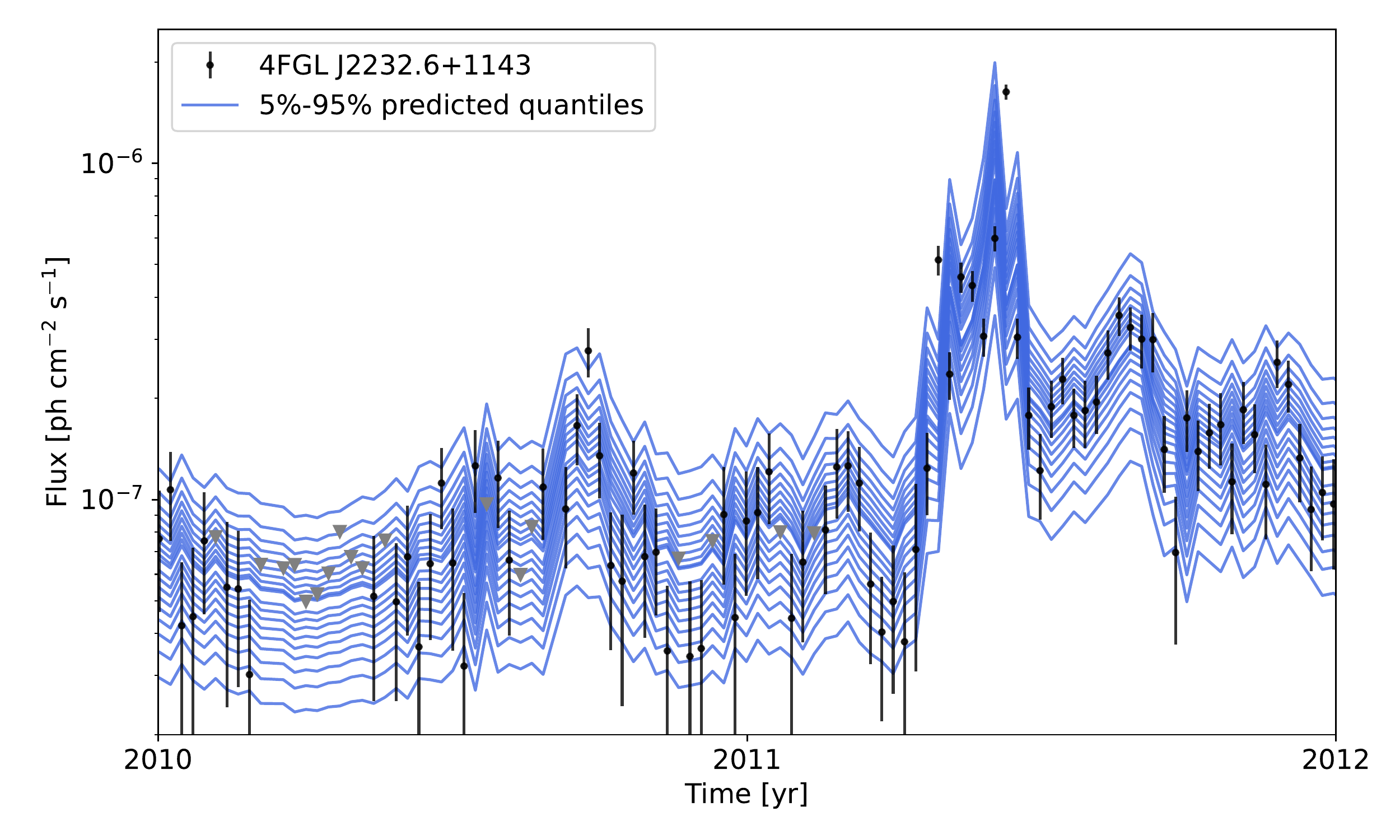}
    \caption{Example of quantiles generated by the trained model, for a portion of the light curve of the most variable LCR source (Fig.~\ref{fig:light_curves}). Gray triangles indicate upper limits.}
    \label{fig:lightcurve_zoom_quantiles}
\end{figure}

An example of the output of the model trained on real data is shown in Fig~\ref{fig:lightcurve_zoom_quantiles}. For each data point of each source, predicted quantiles were generated by using as input the interval of length 101 centered on that data point, replacing the data for that point but no others with a mask encoding. No quantiles were generated for the first and last 50 data points.

The trained models were then used to perform a search for time-reversal asymmetry in flux variability on weekly timescales. We considered the simple case in which the distribution of $\log(F)$ is assumed to be normal with a standard deviation that may vary at different time steps. To do so, a time-reversed set of predicted quantiles was generated as above, except that the time indices within each interval were reversed. The distributions were centered at 0 by subtracting their respective median. The standard deviation of $\log(F)$ was estimated from each set of forward and time-reversed quantiles as the mean of their values divided by the respective values for a normal distribution. For each data point, the relative difference $\Delta_\mathrm{rel}$ between the forward standard deviation $\sigma_\mathrm{for}$ and time-reversed one $\sigma_\mathrm{rev}$ was defined as

\begin{equation}
    \Delta_\mathrm{rel} = 2\frac{\sigma_\mathrm{for} - \sigma_\mathrm{rev}}{\sigma_\mathrm{for} + \sigma_\mathrm{rev}}.
\end{equation}

\begin{figure}[ht]
    \centering
    \includegraphics[width=0.9
    \columnwidth]{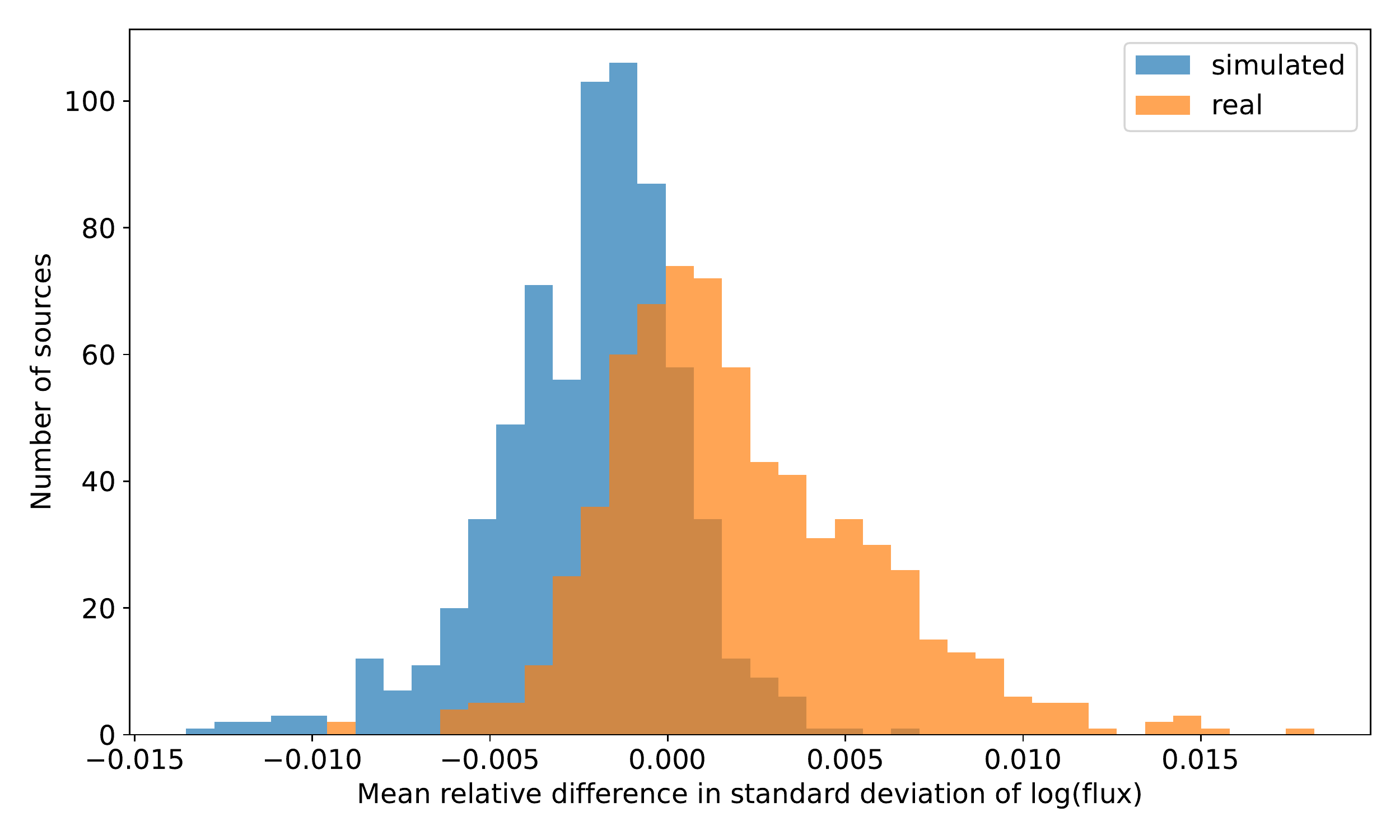}
    \caption{Distributions of the mean relative difference in standard deviation of $\log(F)$ for the simulated and real datasets.}
    \label{fig:time_reversal_histogram}
\end{figure}

The distributions of the mean relative difference $\mu_{\Delta_\mathrm{rel}}$ for each source in the simulated and real datasets are shown in Fig.~\ref{fig:time_reversal_histogram}. The average value of $\mu_{\Delta_\mathrm{rel}}$ for the simulated dataset was $-0.002 \pm 0.003$ and the average for the real dataset was $0.002 \pm 0.004$. No significant deviation from 0 was observed for either dataset. By construction, the mean relative difference for the simulated dataset should be 0 for every source. One possible explanation for a negative relative difference could be the network overfitting on the time orientation it was trained on, causing it to be slightly more confident in the forward direction. For the real data, we therefore adopted 0.002 as an additional systematic uncertainty on the mean relative difference. Incorporating this systematic uncertainty, a preliminary upper bound was placed on the time asymmetry of the amplitude of weekly variability of $\left| \mu_{\Delta_\mathrm{rel}} \right| \lesssim 0.01$ at the 95\% confidence level.

\section{Discussion}

In this work, we investigated self-supervised learning, a potentially powerful new  tool for astrophysical data analysis. As a proof of concept, we used predicted flux distributions generated by a self-supervised Transformer encoder network to perform a novel search for weekly-timescale time-reversal asymmetry in gamma-ray blazar light curves. No significant time asymmetry in the amplitude of the flux distribution was found, suggesting that the amplitude of gamma-ray variability in blazars does not itself vary on weekly timescales. This finding is relevant for many statistical methods commonly used in high-energy astrophysics that assume time-reversal symmetry, such as autoregressive modeling. 

More generally, the self-supervised framework has a number of advantages relative to standard time series analysis methods used with astrophysical data. Upper limits, measurement errors, and missing data can be naturally accommodated using learned encodings. The neural network output is independent of any statistical model or parameterization and can be used for a variety of tasks, making it an unbiased basis for comparison between models or for exploratory searches. In particular, although we did not rely on this property in this work, the Quantile Head generates non-parametric probability distributions. It can therefore be used as a foundation for a variety of studies, such as comparing different intermediate-timescale flux probability distributions. In addition, the trained network encodes knowledge of and generalizes variability patterns learned from an entire dataset, as opposed to a statistical model which must be fit to one light curve at a time. Finally, although we have focused exclusively on gamma-ray data in this work, the network architecture presented here would naturally extend to accommodate data from multiple wavebands at once, potentially allowing novel multiwavelength variability analyses to be conducted.

Multiple avenues exist for future studies to build on the method discussed here. The flux distributions predicted by the network in the present architecture mix together several sources of variability, incorporating not only variability due to physical processes in the blazar but also measurement error; nuisance parameters such as luminosity and redshift; and estimation error due to the finite length of the light curves. One potential way to eliminate the variability caused by measurement error from the predicted distributions would be to add a secondary self-supervised objective to predict the measurement error at each time step. The predicted error distribution could then be deconvolved from the predicted flux distribution using a Monte Carlo or other approach. In addition, physically-motivated data augmentations, such as stretch and scale transformations to mimic the effect of changing redshift, could be applied to help the network disentangle the effects of cosmological nuisance parameters, as well as to enhance the effective size of the dataset. However, because measurement error is nonlinearly correlated with flux, among other factors, doing so would likely require a more detailed model of the \textit{Fermi}-LAT sensitivity. These methods may provide an improved estimation of the intrinsic variability, enabling more detailed studies of the predicted flux probability distributions. Unsupervised clustering and correlation studies using the generated output encodings could also be performed.

\bibliography{aaai23}

\section{Acknowledgments}
A.B. is supported by the NASA Postdoctoral Program at NASA Goddard Space Flight Center, administered by Oak Ridge Associated Universities. Thanks to Daniel Kocevski, Michela Negro, and Janeth Valverde for helpful discussions regarding the \textit{Fermi}-LAT Light Curve Repository. This research has made use of NASA's Astrophysics Data System Bibliographic Services.

\end{document}